\documentclass[11pt,preprint]{aastex}
\usepackage{graphics}
\usepackage{epsfig}

\newcommand{\gsim}{\raisebox{-0.3ex}{\mbox{$\stackrel{>}{_\sim} \,$}}}

\def\xr {X-Ray }

\def\mt {$M-T_{\rm sl}$ }
\begin{document}

\slugcomment{Submitted Version}
\shorttitle{Modeling the observed SZ scaling relations}
\shortauthors{Chaudhuri and Majumdar}

\title{Sunyaev-Zel'dovich scaling relations from a simple phenomenological model
for galaxy clusters}
\author{Anya Chaudhuri and Subhabrata Majumdar}
\affil{Departments of Astronomy \& Astrophysics, Tata Institute of Fundamental
Research, 1, Homi Bhabha Road, Mumbai 400 005}
\email{anya@tifr.res.in}\email{subha@tifr.res.in}

\begin{abstract}

We build a simple, {\it top-down} model for the gas density and temperature
profiles for galaxy clusters. The gas is assumed to be in hydrostatic equilibrium along with a
component of non-thermal pressure taken from simulations and the gas fraction 
approaches the cosmic mean value only at the virial radius or beyond. 
The free parameters of the model are the slope and normalisation of the concentration-mass relation, the gas polytropic index, and slope and normalisation of the mass-temperature relation. These parameters can be fixed from X-Ray and lensing observations.
We compare our gas pressure profiles to the
recently proposed `Universal' pressure profile by \cite{Arnaud09} and find very good agreement.
We find that the Sunyaev-Zel'dovich Effect (SZE) scaling relations between 
 the integrated SZE flux, $Y$, the cluster gas temperature, $T_{\rm sl}$,
the cluster mass, $M_{\rm tot}$, and the gas mass, $M_{\rm gas}$  are in excellent agreement with the recently observed $r_{2500}$ SZE scaling relations by \cite{Bonamente08} and $r_{500}$ relation by \cite{Arnaud09}.
The gas mass fraction increases with cluster mass and is  given by 
$f_{\rm gas}(r_{500}) = 0.1324 + 0.0284 \,\rm{log}\, (\frac{M_{500}}{10^{15}h^{-1}M_\odot})$. This is within $10\%$ of observed 
$f_{\rm gas}(r_{500})$.
The consistency between the global properties of clusters detected in X-Rays
and in SZE shows that we are looking at a common population of clusters as a
whole,  and there is no deficit of SZE flux relative to expectations from \xr scaling 
properties. Thus, it makes it easier to compare and cross-calibrate clusters from
upcoming \xr and SZE surveys.

\end{abstract}								       
									


\section{Introduction} \label{sec:intro}

Large yield SZE cluster surveys promise to do 
precision cosmology once cluster mass-observable scaling relations are reliably
calibrated. This can be done through
cluster observations \citep{Benson04, Bonamente08},  simulations
\citep{Silva04,Bonaldi07} and analytic modeling \citep{Bulbul10}.
It is well known that different astrophysical processes
influence the cluster mass-observable relations non-trivially  (for example, see
\citet{Balogh01,Borgani04, Kravtsov05, Puchwein08}) which can lead to
biases in determining cosmology with clusters.
Alternatively, one can  `self-calibrate' the uncertainties
\citep{ 2003ApJ...585..603M,2004ApJ...613...41M, 2004PhRvD..70d3504L}.

Simplistic modeling of the intra-cluster medium (ICM), like the `isothermal
$\beta$-model' can  give rise to inaccuracies. More complex modeling needs additional
assumptions (such as gas following dark matter at large radii \citep{Komatsu01}, hereafter KS) or inclusion of less understood baryonic physics \citep{Ostriker05}.

To partially circumvent our incomplete knowledge of cluster gas physics, we
build a {\it top-down} phenomenological model of cluster structure, taking clues
from both observations and simulations. 
It stands on three simple, well motivated,
assumptions: (i) present \xr observations can give reliable
cluster mass-temperature relations at $r<r_{500}$ which is used to calibrate our models; (ii) the gas mass fraction, $f_{\rm gas}$, increases with radius as seen in observations \citep {Vikhlinin06, Sun08} and in simulations \citep{Ettori06}, with non-gravitational processes pushing the gas outwards.  It reaches values close to universal baryon fraction  at or beyond the virial radius; and (iii) there is a component
of non-thermal pressure support whose value relative to thermal pressure can be
inferred from biases in mass estimates found in simulations (see 
\cite{Rasia04}). This simple model can reproduce the `Universal' pressure profile \citep{Arnaud09}, \xr gas fraction,  and SZE
scaling relations in excellent agreement with observations.

\section{The Cluster Model}
\subsection{The Cluster Mass Profiles}

The NFW profile \citep{NFW97} is typically used to describe the dark matter mass profile. Here we adopt a NFW form  for the total matter profile since we use the observationally estimated concentration parameter given by \cite{Comerford07} 
$c_{vir} = \frac{14.5 \pm 6.4}{1 + z}( M_{\rm vir}/M_\star)^{-0.15 \pm 0.13}$.
Here $M_\star = 1.3 \times 10^{13} h^{-1} M_\odot$.
The virial radius, $r_{\rm vir}$, is calculated from the spherical collapse model
\citep{Peebles80} as $r_{\rm vir} =
\left[\frac{M_{vir}}{\frac{4\pi}{3}\rho_{crit}(z)\Delta_c(z)}\right] ^{1/3}$.
Here,  $\Delta_c(z) = 18\pi^2 + 82x -39x^2$  \citep{Bryan98}  and
$x = \Omega_m(z) - 1$.

\subsection{The Temperature and density Profiles}
XMM-Newton and Chandra observations have shown that the cluster temperature
declines at large radii \citep{Arnaud05, Vikhlinin06} for both cool (CC) and non-cool core (NCC)
clusters. Simulations \citep{Ascasibar03, Borgani04}, observations \citep{Sanderson2010} and analytic studies \citep{Bulbul10} indicate polytropic profiles for gas temperature. These  studies also point towards an almost constant polytropic index $\gamma \sim 1.2$ (atleast, till $r_{500}$). Hence,  we adopt $T(r) = T(0)f(r)^{\gamma-1}$ and $\rho(r) =\rho(0)f(r)$ . We take the fiducial $\gamma$ = 1.2.

We have also compared the resulting temperature profiles with recent observations \citep{Arnaud07, Sun08} and find the decrements to be comparable . 
Alternatively, for our ``Best-fit''  models we let the $\gamma$ vary along with the $M_0$ and $\alpha$ of the M-T relation and find the values that give the best-fit to the SZ Scaling relations.
Further, CC clusters are characterized by central
temperature decrements which we take to be $T(r)\propto r^{0.3}$  \citep{Sanderson06}
below $0.1 r_{500}$.

To calculate ICM density and temperature profiles, we use the gas dynamical equation \citep{BinneyTremaine, Rasia04}: 
\begin{equation}
\frac{d\Phi(r)}{dr} \,=\, \frac{1}{\rho(r)}\frac{dP(r)}{dr} +
\frac{1}{\rho(r)}\frac{d[\rho(r)\sigma_r^2(r)]}{dr} +
2\beta(r)\frac{\sigma_r^2(r)}{r} \;,
\label{eqn:dynamics}
\end{equation}
where $\Phi$ is the gravitational potential,  $\sigma_r^2(r)$ is the gas velocity
dispersion, 
$\beta(r)$ is the velocity dispersion anisotropy parameter (put equal to zero in
this work) and
$P(r)$  and $\rho(r)$, the gas pressure and density where $ P(r) = \frac{\rho(r)}{\mu m_p}kT(r)$.
Here, $m_p$ is the proton mass and $\mu$ is the mean molecular weight. 
For hydrostatic equilibrium without non-thermal pressure, 
 $\frac{d\Phi(r)}{dr} = \frac{1}{\rho(r)}\frac{dP(r)}{dr}$.
This is normally used to obtain ICM profiles for a given halo
(for example, in \cite{Komatsu01}).  

Simulations \citep{Rasia04, Battaglia10} show that non-thermal pressure can be significant especially at large radii.
Both observations \citep{Mahdavi08, Zhang08}  and simulations \citep{Nagai07, Lau09} suggest that 
the cluster mass calculated assuming only
hydrostatic equilibrium is 
less than the true mass of a cluster.
This discrepancy  increases with radius. Typical values are  20-40\% at $r_{\rm vir}$.
The velocity dispersion term arises from the bulk motions of the ICM and contributes to the non-thermal pressure support.
The profile f(r) can thus be numerically obtained by solving equation 1.

\begin{figure}[ht]
\centering
\includegraphics[width=80mm]{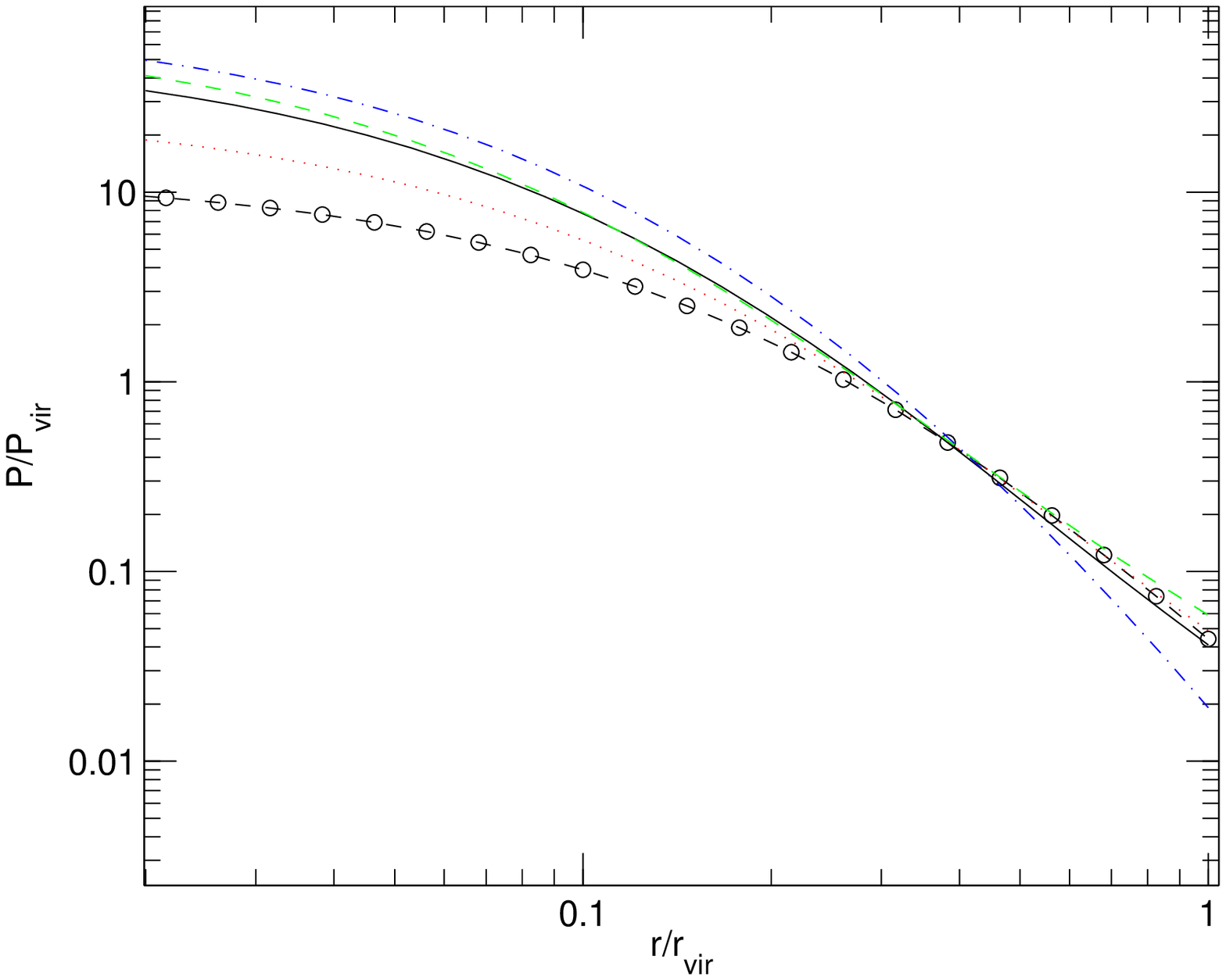}
\caption{ The normalized ICM pressure, $P/P_{vir}$ is plotted against cluster radius for a cluster of mass $5\times10^{14} h^{-1}$ $M_\odot$ and z = 0.  The thick black solid line is the fiducial model; the blue dot-dashed line is for $P_{non-th} = 0$; the green dashed line 
has polytropic index changed to 1.12 from 1.2; the red dotted line is for lower concentration. The KS model pressure is given by the black solid line with circles. Note, that the SZE flux is given by the line-of-sight integral of $P(r)$ over a given cluster area.}
\label{fig:PbyP500}
\end{figure}


\begin{figure*}[ht]
\centering
\begin{tabular}{ll}
\epsfig{file=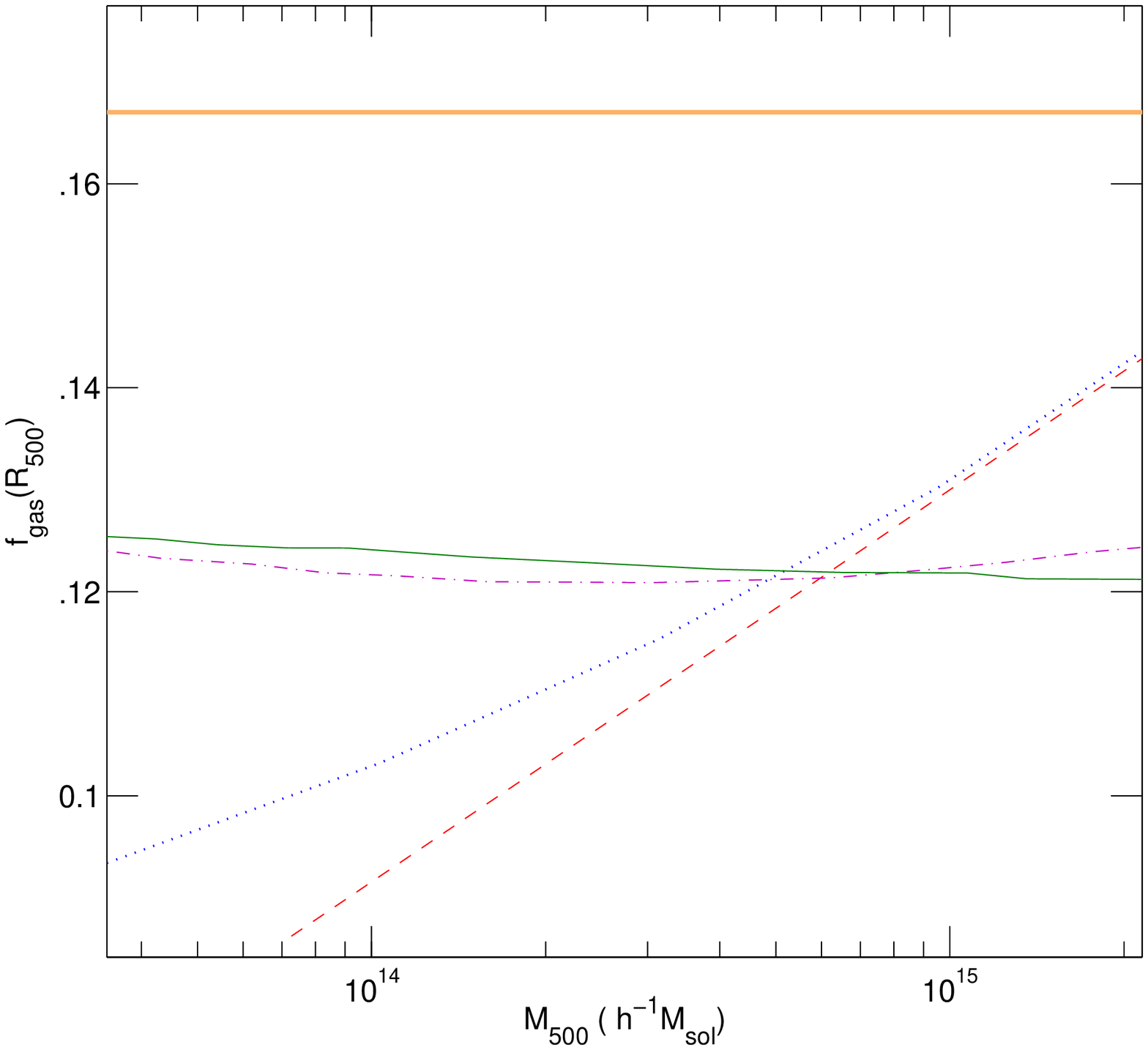,width=.45\linewidth} &
\epsfig{file=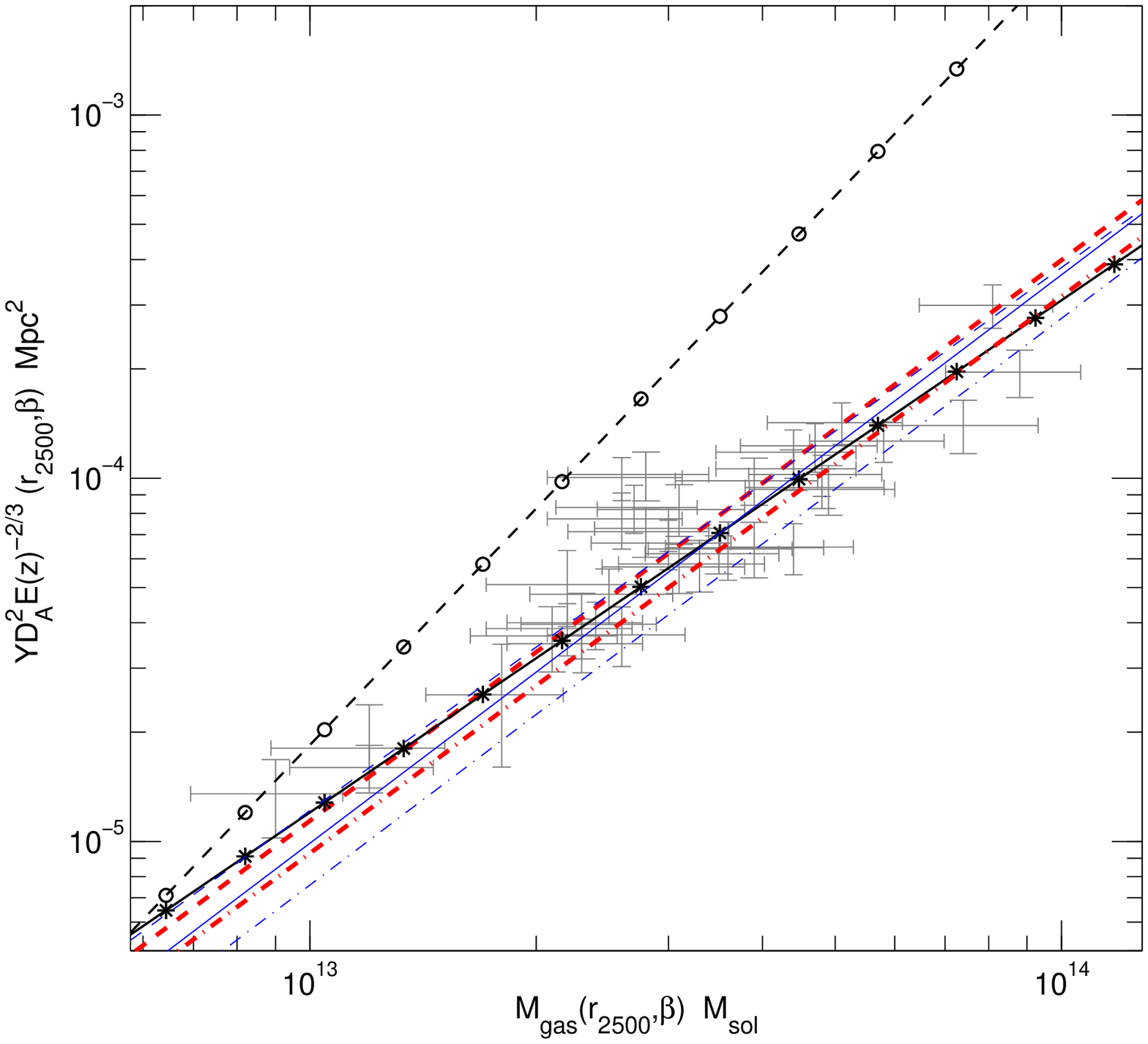,width=.45\linewidth}\\
\end{tabular}
\caption{ Left Panel:  The $f_{gas}$ at $r_{500}$ is plotted against $M_{500}$ for the cluster. The topmost horizontal line is for $f_{\rm gas} = \frac{\Omega_b}{\Omega_m}$;  the solid and dot-dashed lines are for KS model for two different halo concentrations; the dotted line is for our fiducial model and the dashed line is from \xr observations \citep{Vikhlinin09}.  Right Panel: The $Y-M_{\rm gas}$ scaling relations at $r_{2500}$ for our models are shown with the data points from \cite{Bonamente08} and the best fit Bonamente relation.The 5 lines pertaining to our models are - (i) thin blue solid line for model 1 (fiducial), (ii)thin blue dashed line for model 2 , (iii)thin blue dot-dashed line for model 3, (iv) Thick red dot-dashed line  for the model `Bestfit-1' and (v) thick red dashed line for the model `Bestfit-3' (see text for details).  The Bonamente bestfit to data is given by the black solid line with stars. The KS model is shown by dashed line with circles.
}
\label{fig:gas}
\end{figure*}

\subsubsection{Temperature and Density Normalization}
The temperature profiles  are normalized to the recently observed \xr
$M_{500}-T_{\rm sp}$  scaling relation found by \cite{Sun08} which includes data from 
cluster to group scales and is given by
\begin{equation}
 M_{500} E(z) = M_0 \left[\frac{T_{\rm sp}}{3{\rm keV}}\right]^\alpha
\end{equation}
where
$M_0 = (1.21 \pm 0.08)\times  10^{14} h^{-1} M_\odot $ and $\alpha = 1.68\pm .04$.
Here $M_{500}$ is the mass within  $r_{500}$, where  the
average density is $500\rho_c(z)$ where $\rho_c(z)$ is the critical density at redshift z. Using the prescription given by \cite{Mazzotta04} we estimate the
`spectroscopic-like' temperature  $T_{\rm sl}$,  a particular weighted average of
$T(r)$. 
This value of $\alpha \neq 1.5$ indicates deviation from self-similarity, pointing
to non-gravitational energetics in the ICM. 
Here we bypass the microphysics that breaks `self-similarity' but 
normalize the cluster temperatures so as to {\it exactly} reproduce the observed \mt
relation.
Thus our cluster model can be thought of as a {\it top-down} model.

 For any point in parameter space, representative of a simulated cluster, we calculate analytically the temperature and density profiles. We start with an initial arbitrary T(0) and solve for f(r) as described earlier. Next, $T_{\rm sl}$,  is calculated in the radial range $0.1r_{500}$ - $r_{500}$.  The original  T(0) is now adjusted by the ratio of $T_{\rm sl}$ to the $T_{sp}$ from the observed $M_{500}-T_{\rm sp}$ relation. The equation for $f(r)$  is now solved with this new  $T(0)$ after which the $T_{\rm sl}$ is again calculated.  In a few iterations, a self consistent profile f(r) is obtained. Next, 
$\rho(0)$ is determined  by equating the $f_{gas}$ within the cluster radius to 0.9($\Omega_b$/$\Omega_m$) at the cluster boundary ($r_{200}$ or beyond).
The Universal baryon fraction $\Omega_b$/$\Omega_m$ 
is given by 0.167 $\pm$ .009 \citep{Komatsu_WMAP7}.

Simulations \citep{Ettori06} and observations \citep {Vikhlinin06, Sun08} show the gas mass fraction, $f_{gas}(r) =\frac{ M_{\rm gas}(r)}{M_{tot}(r)}$,  
increases with radius. Stellar mass which accounts for a finite fraction of the baryons  is larger at smaller radii such as $r_{2500}$ and for group scale haloes, as observed in the above mentioned studies. Radiative simulations tend to underestimate $f_{gas}$ due to overcooling and predict 
$f_{gas} = 0.7-0.8 (\Omega_b/\Omega_m
)$ at $r_{vir}$. Allowing 10\% of the baryons to form stars, we take $f_{gas}=0.9 (\Omega_b/\Omega_m)$ at the cluster boundary. The resulting $f_{gas}$ as seen in fig \ref{fig:gas}  shows good agreement with the observations at $r_{500}$.
We assume that non-gravitational effects only redistribute the gas.  Recently, both observations \citep{Rasheed10} and theoretical studies \citep{Battaglia10, NathMajum10} show that gas is driven outside the virial radius $r_{vir}$ and atleast upto $2r_{vir}$.

\begin{figure*}[ht]
\centering
\begin{tabular}{ll}
\epsfig{file=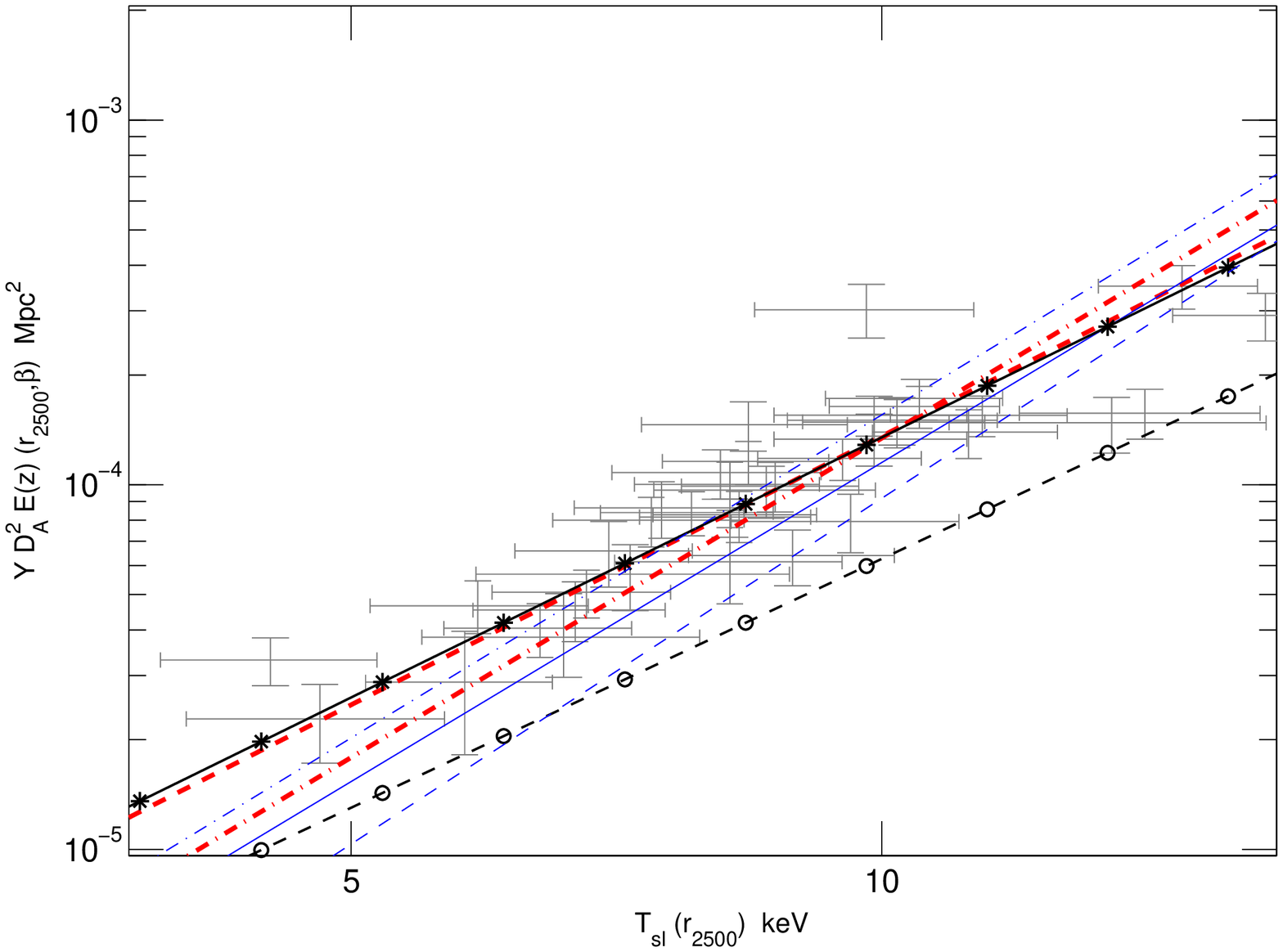,width=.45\linewidth} &
\epsfig{file=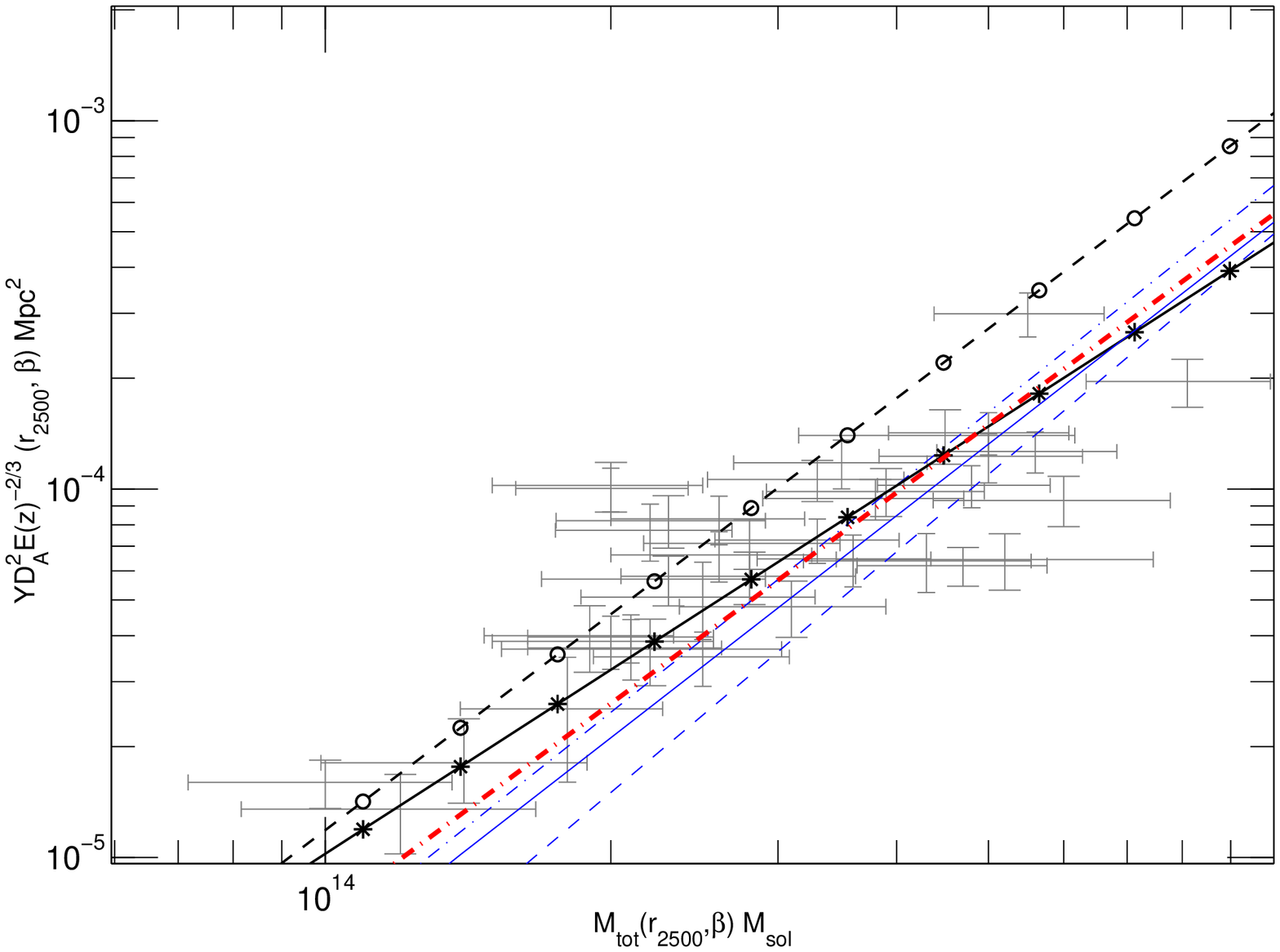,width=.45\linewidth}
\end{tabular}
\caption{Left panel : The $Y-T_{\rm sl}$  relation. The lines here have the same meaning as in figure \ref{fig:gas}, except that the thick red  dashed line is now for our model `Bestfit-2' (see text). Right panel: The $Y-M_{\rm tot}$ relation at $r_{2500}$ is plotted. The line scheme is the same.}
\label{fig:temp_mtot}
\end{figure*}

\begin{table*}[htb]
\begin{footnotesize}
\caption{{SZ Scaling Relations at $r_{2500}$ : ${\rm log} (YD_A^2E(z)^\delta) = A + B {\rm log} (X/c_x)$
where $c_{\rm T_{sl}} = 8{\rm keV},  c_{\rm m_{gas}} = 3*10^{13}M_\odot$ and $c_{\rm m{tot}} =3*10^{14}M_\odot$ and 
$(\delta=1,-2/3,-2/3)$ for $(X=T_{\rm sl}, M_{\rm gas}, M_{\rm tot})$.}}
\begin{tabular}{ l l l l l l l l l l}
\hline
\hline
&\multicolumn{3}{c}{Y-$T_{sl}$}&\multicolumn{3}{c}{Y-$M_{\rm
gas}$}&\multicolumn{3}{c}{Y-$M_{\rm tot}$}\\
\\
\hline
 &   A &  B &  $\Delta\chi^2$ &   A &  B &  $\Delta\chi^2$ &   A &  B &  
$\Delta\chi^2$ \\
\hline
Bonamente & -4.10$\pm$0.22 &  2.37$\pm$0.23 & .0017 & -4.25$\pm$1.77 &
1.41$\pm$0.13 &.073 & -4.20$\pm$3.00 &  1.66$\pm$0.20 & .047\\\\
Best fit & -4.094 &  2.363 &  & -4.25 &
1.414 && -4.19 &  1.654& \\
\hline
\\

model1    &-4.215  &  2.906 & 0.57    & -4.258   & 1.565 &0.12  & -4.322 
 &  2.003 & 0.71\\
model2    &-4.342 &  3.149  & 2.04   & -4.201   & 1.495 & 0.024 & -4.441  &
 2.170& 1.83 \\
model3    &-4.093 &  2.944  &0 .19    & -4.376   & 1.56  &1.31 & -4.245 
 &  2.047 & 0.24 \\
 K-S       &-4.410  & 2.28    & 6.64   & -3.630   & 2.18  &11.55& -4.050
  &1.95    & 0.46 \\           
 \hline
 Bestfit-1 &-4.153 &  2.909  & 0.20    & -4.301   & 1.533& 0.39  &
-4.247   &  1.902& 0.23\\
Bestfit-2      &-4.103&  2.448 &0 .0067    & -   & - & - &
-   &  - &  - \\
Bestfit-3& - &  -  &  -   & -4.207   & 1.544& 0.038  & 
-  &  - &  - \\
\hline
\end{tabular}
\footnotetext{}
\label{table:r2500}
\end{footnotesize}
\end{table*}

\subsection{Model Descriptions}
We include non-thermal pressure,  $P_{\rm non-th}$,  in our calculations. However, this
contribution to the total pressure  ($P_{tot}$) for a
cluster is difficult to model analytically.

In this work, we follow gas dynamical simulations by \citet{Rasia04} to estimate the $P_{\rm non-th}$. We adopt their $P_{th}$/$P_{tot}$ as an input to our model. The mass of the 
cluster calculated from the
hydrostatic term only is lower than the true mass by  $\sim 15\%$ at
$r_{500}$,  $\sim 30\%$ at $r_{vir}$ and $\sim 40\%$ at $2r_{vir}$ in our fiducial model. These values when compared with figure 13 in their paper are found to be of comparable magnitude.

We consider the following  models :

\begin{itemize}
\item model 1 (the fiducial model):  Here $f_{\rm gas} = 0.9
  (\frac{\Omega_b}{\Omega_m})$ at $r = r_{\rm vir}$ ; $M_0 = 1.728\times10^{14}M_\odot$, $\alpha$ = 1.68 and $\gamma$ =1.2. We follow \cite{Rasia04} to estimate $P_{\rm non-th}/P_{\rm total}$.

\item model 2: similar to model 1 but
$f_{\rm gas} = 0.9 (\frac{\Omega_b}{\Omega_m})$ at $r = 2r_{\rm vir}$.  $P_{\rm non-th}/P_{\rm total}$ is extrapolated beyond $r_{vir}$ following simulations by Rasia\footnote{Private communications}.

\item model 3: parameters same as in model 1 but for `zero' $P_{\rm non-th}$.
\end{itemize}
Other than these models, we look at variations of the fiducial model, where we vary the parameters $M_0, 
\alpha \,{\rm and}\, \gamma$, to get the best fit to the \cite{Bonamente08} SZE data. These are called
 Bestfit-1, Bestfit-2 and Bestfit-3 and give a minimum to $\chi^2_{tot} = \chi^2_{(Y-T_{\rm sl})} + \chi^2_{(Y-M_{\rm gas})} + \chi^2_{(Y-M_{\rm tot})}$,  $\chi^2_{(Y-T_{\rm sl})}$, 
  and  $\chi^2_{(Y-M_{\rm gas})}$ respectively, where the $\chi^2$ is to Bonamente data.

 In figure \ref{fig:PbyP500}, we show the effect of varying some of the model parameters on the ICM pressure profile for a $5\times10^{14}h^{-1}M_\odot$ cluster normalized to $P_{vir}$ which is taken to be the ICM pressure at $r_{vir}$ for the standard self-similar model (see \cite{Arnaud09}, Appendix A). Inclusion of non-thermal pressure leads to shallower slope at large radii compared to only thermal pressure. The polytropic index has little influence on the pressure profile for the given change in $\gamma$. The integrated SZE, unlike X-Ray, is similar for both CC and NCC clusters.
In figure \ref{fig:gas} (left panel), we show that clusters in our model naturally have a mass dependent gas fraction, in agreement with observed $f_{\rm gas}$ to within 10\% for 
$M_{500}\gsim2\times10^{14}h^{-1}M_\odot$. For our fiducial model, we find :\\
 $f_{\rm gas}(r_{500}) = 0.1324 + 0.0284 $ $\rm{log} (\frac{M_{500}}{10^{15}h^{-1}M_\odot})$.

\section{The SZE scaling relations - observations and and theoretical models}
The measurement of SZE \citep{Sunyaev80} has come of age in recent times with improvement in detector technologies. Both targeted observations (say from OVRO/BIMA/SZA) and blank sky surveys (ACT/SPT) are underway having much cosmological potential \citep{szARAA02}. Targeted observations have recently given us the SZE scaling relations which can now be used in surveys as proxy for mass.

The  SZE scaling relations  \citep{Bonamente08} predicted from self-similar theory are :
\begin{eqnarray}
YD_A^2 &\propto& f_{gas}T_{\rm sl}^{5/2}E(z)^{-1} \nonumber \\ 
YD_A^2 &\propto& f_{gas}M_{\rm tot}^{5/3} E(z)^{2/3} \nonumber \\
YD_A^2 &\propto& f_{gas}^{-2/3}M_{\rm gas}^{5/3}E(z)^{2/3}
\label{eqn:szscaling}
\end{eqnarray}
where  $Y$ is the the integrated SZE flux from the cluster,
$D_A$ is the angular diameter distance. $M_{\rm gas}$ and $M_{\rm tot}$  are the gas mass and total mass.

\subsection{The $r_{2500}$ Scaling Relations}
\cite{Benson04} presented the first observed SZE scaling relations between the central decrement, $y_0, Y$ and $T_{\rm sl}$ for a sample of 14 clusters. Recently, 
 \cite{Bonamente08} have published scaling relations for 38 clusters at 
$0.14 \leq z \leq 0.89$  using Chandra \xr observations and radio observations with  BIMA / OVRO. Weak lensing mass measurements, at 
$r_{4000-8000}$, of SZE clusters have now been done by \cite{Marrone09} to give the 
$Y-M_{\rm tot}$ scaling. Their extrapolated masses at $r_{2500}$ show agreement to within $20\%$ to the hydrostatic mass estimates by \cite{Bonamente08}.
We follow \cite{Bonamente08} in constructing our scaling relations. In particular we fit beta profiles to the density profile as well as the x-ray surface brightness $S_X$ and compton 
y parameter obtained by projecting the temperature and density profiles obtained as a result of solving equation 1. : 
\begin{eqnarray}
 y = \frac{k_B\sigma_T}{m_ec^2} \int Tn_e \,dl ;  \hspace{2cm} S_X&\propto& \int n_e^2 \,dl 
\end{eqnarray}

The isothermal temperature for each cluster is calculated in the same  radial annulus as theirs. The SZE flux is found by integrating the SZE $\beta$-profiles and the total mass assuming hydrostatic equilibrium is estimated using $M_{\rm tot}(r)=\frac{3\beta k_BT_{\rm sl} r^3}{G\mu m_p (r_c^2 + r^2)}$.  With this prescription, we construct  the  three power law scaling relations given in 
equation \ref{eqn:szscaling}. The coefficients for these scaling relations are specified in table \ref{table:r2500}.  For comparison, we also calculate the SZE scaling relations  from $\beta$-fits to the often used `Komatsu-Seljak' (KS) model \citep{Komatsu01}.  A comparison of the SZE scaling relation for our models, the KS model and the Bonamente data are shown in figures  \ref{fig:gas} and 
\ref{fig:temp_mtot} and table \ref{table:r2500}.

The first point to notice is the good agreement of our model scaling relations with the Bonamente best fit. Especially, for the $Y-M_{\rm gas}$ relation, our models are in excellent agreement with observations. This is relevant as observationally the $Y-M_{\rm gas}$ relation has the least uncertainty. However the assumption of a $\beta$-model for the ICM adds to the uncertainty. For the $Y-T_{\rm sl}$ relation, our estimate of $T_{\rm sl}$ is not accurate for lower temperatures and the agreement with the Bonamente data becomes worse. Especially, for lower $T_{\rm sl}$, our models under predict the SZE flux. Note, that for both of these relations, the KS model line lies outside the data points and hence is a very bad fit to SZE scalings.  Our models also under predicts the SZE flux for masses below $M_{2500} \sim 3\times10^{14} M_\odot$\footnote{The current limiting mass for both SPT and ACT surveys is higher.}.   

Next, we discuss our `Bestfit' models. Once we vary the amplitude and slope of the $M-T_{\rm sl}$ relation and the polytropic index $\gamma$, the best fit values of these parameters obtained are in broad agreement with \xr observations. For example, for `Bestfit-1', where we add the $\chi^2$ from all the three scaling relations, our recovered values are $(M_0, \alpha) = (1.73\times10^{14}M_\odot, 1.7)$ which are within $1-\sigma$ of the \xr values \citep{Sun08}. The best fits are weakly sensitive to the value of $\gamma$; the `Bestfit-1' model prefers a $\gamma=1.14$ which is lower than our fiducial value for $\gamma$.
 
In general, the present data has large error bars and scatter and cannot distinguish between different models (with the exception of the KS model).  However, there are three main points to note: 
(i) $Y-M_{\rm gas}$ is affected more by non-thermal pressure, since its presence influences how much gas can be pushed out.  Our models are within $1-\sigma$ of the Bonamente best fit for $Y-M_{\rm gas}$ while KS model is $> 3-\sigma$ away;  
(ii) At $r_{2500}$, non-thermal pressure has lesser influence on the pressure support. Hence, the `only thermal pressure' model is a good fit to the $Y-M_{\rm tot}$ data, followed by the fiducial model. Here, KS model, with no  non-thermal pressure, is also within $1-\sigma$ to the best fit; 
(iii) Since $T_{\rm sl}$ is found by averaging over an region around $r_{2500}$, it is less influenced by the presence of non-thermal pressure and hence the trend in $Y-T_{\rm sl}$ is similar to the trend in $Y-M_{\rm tot}$. However, KS model with its adiabatic normalization of $T(r)$ is $2-\sigma$ away from  
Bonamente best fit.

\subsection{The $r_{500}$ Scaling Relations obtained from XRay Observations}
\begin{figure}[ht]
\includegraphics[width=80mm]
{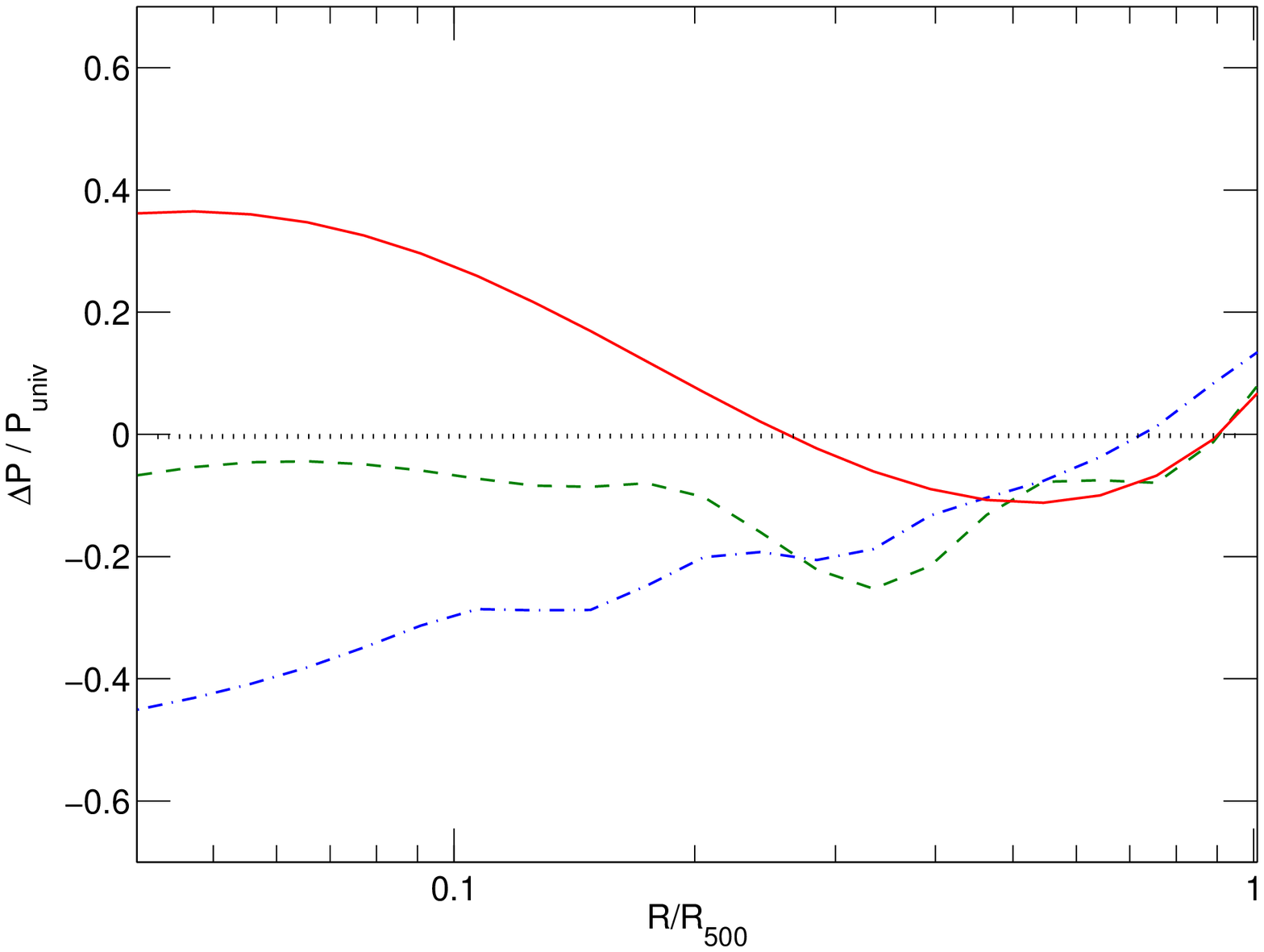}
\caption{Comparison of simulated and semi-analytic gas pressure profiles with observed 'Universal' pressure profile by \cite{Arnaud09} for a cluster having $M_{500} = 2\times10^{14}h^{-1}M_\odot$. Plotted are the fractional differences of our fiducial model given by red solid line, the profiles obtained from simulations by \cite{Battaglia10} (green dashed line) and \cite{Sehgal2010} (blue dot-dashed line) w.r.t the universal  profile found by \cite{Arnaud09} from observations upto $r_{500}$ and simulations beyond.
}
\label{fig:pressure_compare}
\end{figure}

We compare our pressure profiles and scaling relations with the recent 'Universal' pressure profile and resulting SZ scaling obtained by \cite{Arnaud09} from X-Ray observations. We also compare  in fig. \ref{fig:pressure_compare} the pressure profile with those obtained recently by \cite{Battaglia10}, which comes from hydro simulations incorporating a prescription for AGN feedback, and \cite{Sehgal2010} where hot gas distribution within 
halos is calculated using a hydrostatic equilibrium model \citep{Bode09}. Between
 $0.1R_{500}-R_{500}$, i.e.  the core radius and the upper limit for the X-Ray observations, all pressure profiles agree with the observations to within 20\%. All the theoretical pressure profiles start deviating significantly from the observed profile beyond 
$r_{500}$. From the pressure profile, we construct the scaling relation 
$Y_{500}\,=\,{10}^B \left(M_{500}/3 \times10^{14}h_{70}^{-1}\right)^{A} h_{70}^{-5/2}$.  \cite{Arnaud09} find
$B = -4.739 \pm 0.003$ and $A = 1.790 \pm 0.015$. We obtain $(B,A) =(-4.646, 1.670)$  and $(-4.797, 1.805)$  for Model 1 and Model 2 respectively. For the sake of comparison, the values found for \cite{Battaglia10} and 
\cite{Sehgal2010} pressure profiles are $(B,A) = (-4.5\pm0.1, 1.75\pm0.06)$ and $(-4.713\pm0.004, 1.668\pm0.009)$.

\section{Discussions and Conclusion}

We have constructed a {\it top-down} model for galaxy clusters, normalized to the mass-temperature relation from \xr observations.  The gas density and temperature profiles are found by iteratively solving the gas dynamical equation having both thermal and non-thermal pressure support. The form of the  non-thermal pressure used is taken from \cite{Rasia04}. In our model, $f_{\rm gas}$ becomes 0.9 ($\frac{\Omega_b}{\Omega_M}$) at the cluster boundary , whereas gas is pushed  out of the cluster cores to give 
 $f_{\rm gas}(r_{500}) = 0.1324 + 0.0284 $ $\rm{log} (\frac{M_{500}}{10^{15}h^{-1}M_\odot})$,  similar to \xr observations.  

At $r_{2500}$, the SZE scaling relations between SZE flux $Y$ and the cluster average temperature, $T_{\rm sl}$, gas mass, $M_{\rm gas}$, and total mass, $M_{\rm tot}$,  show very good agreement and are within $1-\sigma$ to the best fit line to the \cite{Bonamente08} data. Especially, for the $Y-M_{\rm gas}$ relation the agreement is excellent. In comparison, we also show that the Komatsu-Seljak model is in less agreement to the SZE scaling relations, especially for $Y-T_{\rm sl}$ and $Y-M_{\rm gas}$. Our $r_{2500}$ scaling relations can be compared to those obtained from simulations. For example, the \cite{Nagai06}(see their table 3 for scaling parameters) radiative simulation prediction for the $Y-M_{\rm gas}$ relation gives a $\Delta\chi^2 = 4.7$  w.r.t.  to the best fit \cite{Bonamente08} relation. 
 Recently \cite{Bode09} have predicted SZ scalings from a mixture of N-body simulations plus semi-analytic gas models, normalized to \xr observations for low-$z$ clusters.  The $\Delta\chi^2$ of their model is 0.06 and agrees well with our results.

Further out, at $r_{500}$, the $Y-M$ scaling relation obtained for our models agree very well with those obtained from X-Ray observations \citep{Arnaud09}. Most assuringly, the gas pressure profile in our simple phenomenological model of clusters, comes out to be within $\sim20\%$ beyond .1 $R_{500}$ to the observed `Universal' pressure profile given by \cite{Arnaud09}. 


Most importantly, the fact that \xr normalised models can reproduce SZE scaling relations  well is reassuring for cluster studies. It shows that we are looking at a common population of clusters as a
whole,  and there is no deficit of SZE flux relative to expectations from \xr scaling 
properties. Thus, one can compare and cross-calibrate clusters from
upcoming \xr and SZE surveys with increased confidence. It also gives us confidence to extrapolate our models to larger radii in order to construct the $Y-M_{200}$ scaling relation and SZ power spectrum templates. 

\section*{Acknowledgements}
SM would like to thank Kancheng Li for being a great summer student who started this project, albeit, with a different focus.  The authors thank Jon Sievers, Elena Rasia and Nick Battaglia. SM also wishes to thank Dick Bond and Christoph Pfrommer for many lively discussions on 'gastrophysics'.
 The authors would like to thank the Referee for the valuable comments and suggestions.


\end{document}